\documentclass[12pt]{iopart}
\usepackage[latin1]{inputenc}
\usepackage{graphicx}
\usepackage{iopams}
\usepackage{cite}

\def\be{\begin{equation}}
\def\ba{\begin{eqnarray}}

\def\b{\beta}

\def\D{\Delta}

\def\r{\rho}

\def\i{\int}

\def\Tr{\mbox{Tr}}

\def\ee#1{\label{#1}\end{equation}}
\def\ea#1{\label{#1}\end{eqnarray}}

\begin{document}
\note[The Tasaki-Crooks quantum fluctuation theorem]{The
Tasaki-Crooks
  quantum fluctuation theorem}
\author{Peter Talkner\footnote[7]{Corresponding author:
peter.talkner@physik.uni-augsburg.de},
Peter H\"anggi}

\address{Institut f\"ur Physik, Universit\"at Augsburg,
Universit\"atsstrasse 1, D-86135 Augsburg, Germany}

\begin{abstract}
Starting out from the recently established quantum correlation
function expression of the characteristic function for the work
performed by a force protocol on the system [cond-mat/0703213] 
the quantum version of the Crooks fluctuation theorem
is shown to emerge almost immediately by the mere application of an
inverse Fourier transformation.

\end{abstract}

\pacs{05.40.-a, 87.16.-b, 87.19.Nn}
\submitto{J. Phys. A}


Work and fluctuation theorems have ignited much excitement during the
recent decade \cite{Ev,GC,Cr,Ja}. These theorems have prompted
further theoretical investigations \cite{Ku,T00,Mu,Se} as well as
experimental research \cite{Bu}. We here consider a quantum system
staying in {\it weak thermal contact} with a heat bath at the
inverse temperature $\b$ until a time $t_0$.  At  time $t_0$ the
contact to the heat bath is then either kept at this  weak level, or
may even be switched off altogether. A classical time dependent
force solely acts on the system according to a prescribed protocol
until time $t_f$. A {\it protocol} defines a family of Hamiltonians
$\{H(t)\}_{t_f,t_0}$ which govern the time evolution of the system
during the indicated interval of time $[t_0,t_f]$ in the presence of
the external force. The weak action of the heat bath on the system 
can be neglected 
for any protocol of finite duration $t_f-t_0$ \cite{Sp}. 
The work performed by the force on the system
is a random quantity because of the quantum nature of the considered
system and because the system is prepared in the thermal equilibrium
state \be \r(t_0) = Z(t_0) \exp \{-\b H(t_0)\} \ee{r0} which is a
mixed state for all finite $\b$. Here, $Z(t_0) = \Tr \exp \{-\b
H(t_0)\} $ denotes the partition function. As a random quantity, the
work is characterized by a probability density $p_{t_f,t_0}(w)$ or
equivalently by the corresponding characteristic function
$G_{t_f,t_0}(u)$, which is defined as the Fourier transform of the
probability density, i.e. \be
 G_{t_f,t_0}(u) = \i dw\: e^{iuw} p_{t_f,t_0}(w).
\ee{Gw} In a recent work \cite{t07} we have  demonstrated that the
characteristic function $G_{t_f,t_0}(u)$ of the work can be
expressed as quantum correlation function of the two exponential
operators $\exp\{iuH(t_f)\}$
  and $\exp\{-iuH(t_0)\}$. It explicitly reads:
\be
\eqalign{
G_{t_f,t_0}(u)
&= \langle e^{i u H(t_f)} e^{-i u H(t_0)} \rangle_{t_0}\\
&\equiv Z^{-1}(t_0)\Tr\: U^+_{t_f,t_0} e^{iu
  H(t_f)}U_{t_f,t_0} e^{-iuH(t_0)} e^{-\b H(t_0)} \;,
} \ee{G} where the index at the bracket signifies the fact  that the
average is taken over the initial density matrix $\r(t_0)$.

For a protocol
consisting of Hamiltonians $H(t)$, each of which is bounded from below
and has a purely
discrete spectrum,
the characteristic function $G_{t_f,t_0}(u)$ is an analytic function
of $u$ in the strip
$S=\{ u| 0\leq \Im u \leq \beta, -\infty < \Re u < \infty
\}$~\cite{n1} where $\Re u$ and $\Im u$ denote the real and imaginary
part of $u$, respectively.
Collecting the two exponential factors $e^{-iu
  H(t_0)}$ and $e^{-\b H(t_0)}$ into one, and introducing the complex
parameter $v= -u +i \b \in S$ we find
\be
\eqalign{
Z(t_0) G_{t_f,t_0}(u) &= \Tr\: U^+_{t_f,t_0}\: e^{i(-v+i\b)
  H(t_f)}\:U_{t_f,t_0}\:e^{iv H(t_0)} \\
&= \Tr\: e^{-iv H(t_f)} \:e^{-\b H(t_f)}\: U_{t_f,t_0} \:e^{i v H(t_0)
}\:U^+_{t_f,t_0} \\
&= \Tr \:e^{-iv H(t_f)} \:e^{-\b H(t_f)}\: U^+_{t_0,t_f}\: e^{i v H(t_0)
}\:U_{t_0,t_f} \\
&= \Tr \: U^+_{t_0,t_f}\: e^{i v H(t_0)}\:U_{t_0,t_f}\:e^{-iv H(t_f)}\: e^{-\b
  H(t_f)} \\
&=Z(t_f)\:G_{t_0,t_f}(v)
}
\ee{Guv}
where we used the unitarity of the time evolution operator,
i.e. $U^+_{t_f,t_0} = U^{-1}_{t_f,t_0} = U_{t_0,t_f}$. We hence obtain
\be
 G_{t_f,t_0}(u) = \frac{Z(t_f)}{Z(t_0)} G_{t_0,t_f}(-u+i\b).
\ee{FTG} The ratio of the canonical partition functions can be
expressed in terms of the difference of free energies $\D F$ between
the two thermal equilibrium systems as $Z(t_f)/Z(t_0) = \exp \{ - \b
\D F \}$. The quantity $G_{t_0,t_f}(v)$ coincides with the
characteristic function of the work performed on a system that is
initially prepared in the thermal equilibrium state $Z(t_f)^{-1}
\exp\{-\b H(t_f)$ under the influence of the {\it time-reversed}
protocol $\{ H(t) \}_{t_0.t_f}$. Applying the inverse Fourier
transform on both sides of eq.~(\ref{FTG}) we obtain the following
fluctuation theorem
\be
\frac{p_{t_f,t_0}(w)}{p_{t_0,t_f}(-w)} = \frac{Z(t_f)}{Z(t_0)} e^{\b
  w} = e^{-\b( \D F - w)} \;.
\ee{FTp} It relates the probability density of performed work for a
given protocol to that of the work for the time-reversed  process.
This process can in principle be realized by preparing the  Gibbs
state $Z^{-1}(t_f) \exp \{-\b H(t_f) \}$ as the {\it initial}
density matrix and letting run the time-reversed protocol
$\{H(t)\}_{t_0,t_f}$.

In the classical context this fluctuation theorem was proved by
Gavin Crooks~\cite{Cr}, its quantum version goes back to Hal
Tasaki~\cite{T00}.\\

{\bf Acknowledgments}. This work has been supported by the Deutsche
For\-schungs\-ge\-mein\-schaft via the Collaborative Research Centre
 SFB-486, project A10.  Financial support of the German Excellence Initiative
 via the {\it Nanosystems Initiative Munich} (NIM) is gratefully acknowledged as well.


\begin{thebibliography}{99}
\bibitem{n1} This can be  proved in the same way as the 
analyticity properties of
  equilibrium correlation functions
  that underly the KMS condition,
  cf.~\cite{HHW}.
\end{thebibliography}
\end{document}